\documentstyle[aps,prl,latexsym,epsfig,multicol,times]{revtex}
\makeatletter
\newlength\abovecaptionskip \newlength\belowcaptionskip
\def\@makecaption#1#2{%
 \vskip\abovecaptionskip \sbox\@tempboxa{#1: #2}%
 \ifdim \wd\@tempboxa >\hsize #1: #2\par \else \global \@minipagefalse
 \hb@xt@\hsize{\hfil\box\@tempboxa\hfil}%
 \fi \vskip\belowcaptionskip} \makeatother

\newcommand{\var}{\langle \epsilon^2 \rangle}

\newcommand{\mvar}{\langle \tilde m^2 \rangle}
\newcommand{\mav}{\langle m \rangle}
\newcommand{\eav}{\langle \eps \rangle}
\newcommand{\eps}{\epsilon}
\newcommand{\ve}{\vert}
\newcommand{\fr}{\frac}
\newcommand{\ran}{\rangle}
\newcommand{\lan}{\langle}
\newcommand{\beq}{\begin{equation}}
\newcommand{\eeq}{\end{equation}}

\begin{document}            

\title{A renormalization approach for the 2D Anderson model at the band
edge: Scaling of the localization volume}
\author{Stefanie  Russ\\
Institut f\"ur Theoretische Physik III, Universit\"at Giessen,
D-35392 Giessen, Germany \\
}
 
\maketitle

\begin{multicols}{2}[%
\begin{abstract} 
We study the localization volumes $V$ (participation ratio) of 
electronic wave functions in the $2d$-Anderson model with diagonal disorder.
Using a renormalization procedure, we show that at the band edges, 
i.e. for energies $E\approx \pm 4$, $V$ is
inversely proportional to the variance $\var$ of the site potentials. Using
scaling arguments, we show that in the neighborhood of $E=\pm 4$, $V$ 
scales as $V=\var^{-1}g((4-\ve E\ve)/\var)$ with the scaling function
$g(x)$. Numerical simulations confirm this scaling ansatz.
\end{abstract}
\pacs{PACS numbers: 
71.23.An,  
72.15.Rn,  
63.20.Pw   
}]

\section{Introduction}
A large amount of work has been done in the past decades to understand the
localization behavior in randomly disordered systems. The standard model for
a single-particle electronic wave function in tight binding 
approximation in the presence of disorder is the Anderson model 
\cite{kramer,janssen,anderson}. In $d=2$ and with diagonal disorder, it can be 
written as
\begin{eqnarray} \label{anderson}
&& \psi_{n+1,m} + \psi_{n-1,m} + \psi_{n,m+1} + \psi_{n,m-1} - 
4 \psi_{n,m} \nonumber\\
&& =  (E-4) \psi_{n,m} - \eps_{n,m} \psi_{n,m}.
\end{eqnarray}
Here $E$ is the energy, the hopping potentials between nearest-neighbor
are all set to unity, $(n,m)$ are the site 
indices, $\psi$ is the eigenfunction, and $\ve\psi_{n,m}\ve^2$ is the 
probability to find an
electron at site $(n,m)$. The $\eps_{n,m}$ are the site potentials which are  
all uncorrelated random numbers with the variance $\var \equiv (1/N^2)
\sum_{n,m=1}^{N} \eps_{n,m}^2$.
Their average value $\langle\eps\rangle\equiv (1/N^2)\sum_{n,m=1}^{N} 
\eps_{n,m}$ is set equal to zero. The term $-4\psi_{n,m}$ on both sides of
Eq.~(\ref{anderson}) is introduced in order to create a
discretized Laplace operator on the left-hand side of the equation (see below).
 
It has been recognized for long that in $d=1$ and $d=2$ all eigenstates 
of Eq.~(1) are 
localized, whereas a localization-delocalization transition occurs in $d=3$. 
However, the shape of the wavefunctions and the value of the localization
length $\lambda(E,\var)$ is still being discussed. 

In $d=1$ and for uncorrelated site potentials, exponential localization was 
proven throughout the energy band \cite{kramer,borland,halperin} and a lot of 
rigorous results and scaling theories exist for the localization length $\lambda$,
defined via the Lyapunov exponent.
Close to the band edges (i.e. at $E=\pm 2$ in $d=1$), 
a weak disorder expansion yields 
\beq\label{lok1}
\lambda = \var^{-\alpha}f\left(\frac{E_c-\vert E\vert}{\var^{\beta}}\right)
\eeq 
with $E_c=2$, $\alpha=1/3$ and $\beta=2/3$ \cite{gid,dergard}. 
Recently, it has been shown by a space renormalization 
procedure \cite{philmag,russprb} that Eq.~(\ref{lok1}) also holds for
the case of long-range correlated site potentials with correlation exponent
$\gamma$, $0<\gamma\le 1$. In this case, the
exponents have to be replaced by $\alpha=1/(4-\gamma)$ and $\beta=2/(4-\gamma)$
and $\gamma=1$ refers to the uncorrelated case of Refs.~\cite{gid,dergard}.
At the band center on the other hand, a different behavior of $\lambda$ occurs. 
A Green's function technique \cite{kappus} yields $\lambda\sim \var^{-1}$ for $E=0$ and 
in some distance from the band center, a second-order perturbation 
theory of the diagonal elements of the Green's
function \cite{kramer,thouless} yields $\lambda(E) \sim (4-E^2)/\var$.

For $d=2$, on the other hand, no analytical theory for the localization behavior
is known yet. 
Numerical simulations close to the band center exist on the
basis of Green's functions calculations \cite{kinnon81}, exact diagonalization 
\cite{zarek} and the Lanczos algorithm \cite{kabu2002}, but do not lead
to an exact or scaling form of $\lambda$ or related quantities. 
Moreover, it was shown 
in Ref.~\cite{kabu2002}, that the wave function in the two-dimensional
Anderson model does not decay exponentially. Instead, a subexponential
decay of $\psi$ was found with $\lambda$ increasing logarithmically with 
the distance $r$ from the localization center.

In this paper, we concentrate on the band edges, i.e. on energies 
$E\approx\pm 4$ of the Anderson model in $d=2$ with uncorrelated potentials. 
We develop a renormalization 
approach, similar to the one in $d=1$ of \cite{philmag,russprb} and use it 
to find a scaling form for the localization volume $V$, which is related to the
inverse participation ratio $P^{-1}$. In $d=2$ and with the wave function 
$\psi_{n,m}$ being normalized by $\sum_{n,m}\psi^2_{n,m}=1$, $P^{-1}$ is defined 
by \cite{wegner,thouless74}
\beq \label{ipr}       
P^{-1} = \sum_{n,m=1}^{N} \ve\psi_{n,m}\ve^4.       
\eeq       
Its inverse value $P$ is a $d$-dimensional volume and
measures the extension of a given state. If we divide $P$ by the volume $V_0$ 
of the system, we get the relative volume $V$ of the eigenstate, $V\equiv P/V_0$,
i.e. the portion of the system where the wave amplitude is large. 
It can be easily verified that in $d=1,2$ or $3$, $V\sim\lambda^d$ for all wavefunctions
of the form $\psi(r)\sim\exp[-(r/\lambda)^\Phi]$, $\Phi>0$. Therefore,
one can define an effective localization length 
$V^{1/d}\sim\lambda$, which measures the average diameter of the state.
For numerical calculations in $d=2$ and $d=3$, where the wave functions do not decay
exponentially, $V$ is easier accessible than $\lambda$ and therefore, we focuse on
$V$ in this paper. 

Since $V\sim\lambda$ in $d=1$, Eq.~(\ref{lok1}) holds up to an
irrelevant proportionality factor also for $V$. 
It is the purpose of this paper to show that a similar scaling law as 
the one of Eq.~(\ref{lok1}) holds also for $V$ in $d=2$, 
\beq\label{lok2}
V = \var^{-\alpha}g\left(\frac{E_c-\vert E\vert}{\var^{\beta}}\right)
\eeq 
but with different exponents, $\alpha=1$ and $\beta=1$ and with $E_c=4$.
This scaling ansatz is confirmed by numerical calculations.  

The paper is organized as follows: In section II we explain the outline of the
renormalization approach, while in section III the scaling ansatz for the
localization volume $V$ is developed and tested by numerical
simulations.  Additional remarks about former 
renormalization theories and the extension to the vibrational problem 
are given in section IV. 

\section{The renormalization approach}
In the renormalization approach, we want to combine single sites of
the lattice to blocks. This procedure must be reasonable in the limit of small 
values of $4-\ve E\ve$, i.e. close to the band edges. In this context,
we must recall that the wavefunctions possess two characteristic lengthscales, 
(i) the wavelength $\Lambda$ of the ordered lattice that describes the periodic
fluctuating part of the wavefunction and (ii) the localization length 
that describes their decaying envelope. As in $d=1$ \cite{russprb}, we
assume that the periodic part of $\psi$ does not depend on the disorder but is
reminiscent of the functions of an ordered lattice, where 
the disorder terms $\eps_{i,j}$ are zero and $\psi$ is a regular $\sin$- or 
$\cos$-function. By solving Eq.~(\ref{anderson}) for the ordered case, we found
in $d=1$ that $\Lambda^2\sim(\ve E_c\ve-E)^{-1}$. In $d=2$, where
$\Lambda^2=\Lambda_x^2+\Lambda_y^2$ with the wavelengths $\Lambda_x$ and
$\Lambda_y$ in $x-$ and $y-$ direction, respectively,
we have
\beq \label{wavelenght}
\Lambda_x^2\sim (E_c-\ve E\ve)^{-1}\qquad\mbox{and}\qquad
\Lambda_y^2\sim (E_c-\ve E\ve)^{-1}
\eeq
where $E_c=4$.
At the band edges, $\Lambda$ diverges and the wavefunction no
longer resolves the details of the disorder potentials. In this case, 
we can imagine that neighboring sites of the lattice move as blocks and
the following renormalization approach becomes legitimate. 

In the following, we consider the upper band edge $E = 4$, but by canceling
the terms $-4\psi_{n,m}$ on both sides of Eq.~(\ref{anderson}) and taking into 
account
that the $\eps_{n,m}$ are randomly distributed around their mean value of
$\eav=0$, we can see that the equation is symmetric under the transformation 
$E\to -E$. Therefore, the renormalization approach is also valid for the lower
band edge $E= -4$.

In order to transform Eq.~(\ref{anderson}) into block form,
we first replace the site indices $(n,m)$ of the central site in
Eq.~(\ref{anderson}) (see also Fig.~\ref{bi:sketch1}) successively by
$(n+1,m)$, $(n-1,m)$, $(n,m+1)$ and $(n,m-1)$. Combining those four equations with
Eq.~(\ref{anderson}) and rearranging the terms, we arrive at 
\begin{eqnarray} \label{anderson2}
&& \psi_{n+2,m} + \psi_{n-2,m} + \psi_{n,m+2} + \psi_{n,m-2} - 
4 \psi_{n,m} \nonumber\\
&=&  -\left(4f_{n,m} + f_{n,m+1} + f_{n,m-1} + f_{n+1,m} + f_{n-1,m} \right)
\nonumber\\
&+& (E-4) (4\psi_{n,m} + \psi_{n+1,m} + \psi_{n-1,m} + \nonumber\\
&& \qquad\qquad\qquad +\psi_{n,m+1} + \psi_{n,m-1}) 
+ 8\psi_{n,m} \nonumber\\ 
&-& 2(\psi_{n+1,m+1} + \psi_{n+1,m-1} + \psi_{n-1,m+1} + \psi_{n-1,m-1}),
\nonumber\\
&
\end{eqnarray}
with the abbreviation $f_{i,j}\equiv\eps_{i,j}\,\psi_{i,j}$.
Comparing this result with Eq.~(\ref{anderson}), we can see that
the left-hand side of Eq.~(\ref{anderson2}) is again a Laplace 
operator, but with twice the distance between $\psi_{n,m}$ and its neighbors.
The first two terms on the right-hand side, involving the disorder terms
$f_{i,j}$ and the eigenvalue $(4-E)$, are similar to the corresponding terms in
Eq.~(\ref{anderson}), with the only difference that they no longer depend on 
a single site $(n,m)$ but couple sites at distances $<2$ from $(n,m)$ to blocks. 

The last two terms, however, involve couplings between $\psi_{n,m}$ and its 
second nearest neighbors $\psi_{n+1,m+1}$, $\psi_{n+1,m-1}$ and so 
on and do not occur in Eq.~(\ref{anderson}) (nor in the corresponding 
derivation in $d=1$). 
Using a second-order Taylor expansion, we approximate these terms by
\begin{eqnarray}\label{taylor}
&&\psi_{n+1,m+1}+\psi_{n+1,m-1}+\psi_{n-1,m+1}+\psi_{n-1,m-1}\nonumber\\
&\approx&-4 \psi_{n,m}+2(\psi_{n+1,m}+\psi_{n-1,m}+\psi_{n,m+1}+\psi_{n,m-1})\nonumber\\
&=& 4\psi_{n,m} - 2f_{n,m} + 2(E-4)\psi_{n,m},
\end{eqnarray}
where Eq.~(\ref{anderson}) has been inserted in the last step.
The Taylor expansion is legitimate in the limit of large wavelengths,
i.e. close to the band edge. 
Inserting Eq.~(\ref{taylor}) into Eq.~(\ref{anderson2}) we finally arrive at
\begin{eqnarray} \label{anderson3}
&& \psi_{n+2,m} + \psi_{n-2,m} + \psi_{n,m+2} + \psi_{n,m-2} - 
4 \psi_{n,m} \nonumber\\
&=& - (f_{n,m+1} + f_{n,m-1} + f_{n+1,m} + f_{n-1,m})\nonumber\\
&+& (E-4)(\psi_{n,m+1} + \psi_{n,m-1} + \psi_{n+1,m} + \psi_{n-1,m}).
\end{eqnarray}
Assuming that the potentials are randomly distributed, we
introduce the smoothed 
wavefunction $\psi_{n,m}^{(2)}$ of the block and combine
the terms $f_{n,m+1} + f_{n,m-1} + f_{n+1,m} + f_{n-1,m}$ 
to one single term $f^{(2)}_{n,m}\equiv\eps_{n,m}^{(2)}\psi_{n,m}^{(2)}$ with the 
block potential $\eps_{n,m}^{(2)}\equiv\eps_{n,m+1}+ \eps_{n,m-1}
+ \eps_{n+1,m} + \eps_{n-1,m}$.
Equation~(\ref{anderson3}) now shows a block form of block length $\nu=2$, 
\begin{eqnarray} \label{anderson4}
&& \psi^{(2)}_{n+1,m} + \psi^{(2)}_{n-1,m} + \psi^{(2)}_{n,m+1} + 
\psi^{(2)}_{n,m-1} - 4 \psi^{(2)}_{n,m} \nonumber\\
&&= - \eps^{(2)}_{n,m}\psi^{(2)}_{n,m} + 4(E-4)\psi^{(2)}_{n,m}.
\end{eqnarray}

This is shown in Fig.~\ref{bi:sketch1}. 
Couplings between nearest-neighbor sites via Eq.~(\ref{anderson}) 
are symbolized by straight lines whereas the couplings
between the sites $\psi_{n+2,m}$ and $\psi_{n,m}$ and so on of
Eq.~(\ref{anderson4}) are symbolized by the oval lines. The site potentials 
$f_{n,m+i}\equiv\eps_{n,m+i}\psi_{n,m+i}$ and $f_{n+i,m}\equiv\eps_{n+i,m}
\psi_{n+i,m}$ with $i=\pm 1$ that form the block potential $\eps_{n,m}^{(2)}$
are indicated by the black circles. It can be seen that
they lie well inside an inclined block, consisting of $2^2$ particles.

\unitlength 1.85mm
\vspace*{0mm}
{
\begin{figure}
\begin{picture}(40,30)
\def\epsfsize#1#2{0.4#1}
\put(0,0){\unitlength0.3mm\unitlength0.3mm 
\linethickness{0.4pt} 
\put(30,100){\line(1,0){21}}
\put(69,100){\line(1,0){21}}\put(110,100){\line(1,0){21}}
\put(149,100){\line(1,0){21}}

\put(70,140){\line(1,0){21}}\put(109,140){\line(1,0){21}}
\put(70,60){\line(1,0){21}}\put(109,60){\line(1,0){21}}

\put(100,30){\line(0,1){7}}\put(100,48){\line(0,1){4}}
\put(100,69){\line(0,1){7}}\put(100,87){\line(0,1){4}}
\put(100,110){\line(0,1){7}}\put(100,128){\line(0,1){4}}
\put(100,149){\line(0,1){7}}\put(100,167){\line(0,1){4}}

\put(60,70){\line(0,1){7}}\put(60,88){\line(0,1){4}}
\put(60,109){\line(0,1){7}}\put(60,127){\line(0,1){4}}
\put(140,70){\line(0,1){7}}\put(140,88){\line(0,1){4}}
\put(140,109){\line(0,1){7}}\put(140,127){\line(0,1){4}}

\put(100,100){\circle{20}}
\put(140,100){\circle*{20}}\put(180,100){\circle{20}}
\put(60,100){\circle*{20}}\put(20,100){\circle{20}}

\put(100,140){\circle*{20}}\put(100,180){\circle{20}}
\put(100,60){\circle*{20}}\put(100,20){\circle{20}}

\put(140,140){\circle{20}}\put(140,60){\circle{20}}
\put(60,140){\circle{20}}\put(60,60){\circle{20}}

\put(100,80){\makebox(0,0)[b]{n,m}}
\put(142,80){\makebox(0,0)[b]{n+1,m}}\put(187,80){\makebox(0,0)[b]{n+2,m}}
\put(58,80){\makebox(0,0)[b]{n-1,m}}\put(18,80){\makebox(0,0)[b]{n-2,m}}

\put(50,120){\makebox(0,0)[b]{n-1,m+1}}\put(100,120){\makebox(0,0)[b]{n,m+1}}
\put(150,120){\makebox(0,0)[b]{n+1,m+1}}
\put(50,40){\makebox(0,0)[b]{n-1,m-1}}\put(100,40){\makebox(0,0)[b]{n,m-1}}
\put(150,40){\makebox(0,0)[b]{n+1,m-1}}

\put(100,0){\makebox(0,0)[b]{n,m-2}}\put(100,160){\makebox(0,0)[b]{n,m+2}}

\put(140,100){\oval(80,20)[t]}\put(60,100){\oval(80,20)[t]}
\put(100,60){\oval(30,80)[r]}\put(100,140){\oval(30,80)[r]}}      
\end{picture}
\caption[]{\small Sketch of the Anderson lattice according to 
Eqs.~(\ref{anderson})
und (\ref{anderson3}) as explained in the text. The circles represent the
different lattice sites, the straight lines between them indicate the usual
nearest-neighbor coupling, whereas the couplings of Eq.~(\ref{anderson3})
are shown by the oval lines. The black circles stand for the site potentials that
form the block potential [see Eq.~(\ref{anderson4})].
}
\label{bi:sketch1}
\end{figure}
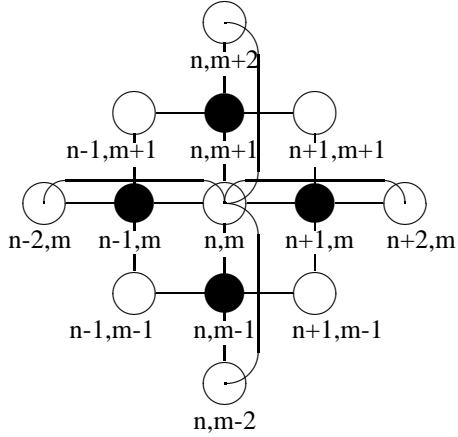}

\unitlength 1.85mm
\vspace*{0mm}
{
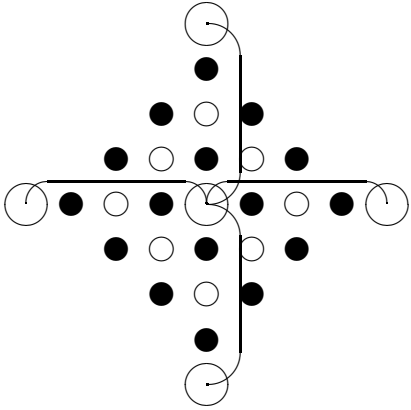
\begin{figure}
\begin{picture}(40,30)
\def\epsfsize#1#2{0.4#1}
\put(0,-2){\unitlength0.3mm\unitlength0.3mm 
\linethickness{0.4pt} 
\put(100,100){\circle{20}}
\put(120,100){\circle*{10}}\put(140,100){\circle{10}}\put(160,100){\circle*{10}}
\put(80,100){\circle*{10}}\put(60,100){\circle{10}}\put(40,100){\circle*{10}}

\put(100,120){\circle*{10}}
\put(120,120){\circle{10}}\put(140,120){\circle*{10}}
\put(80,120){\circle{10}}\put(60,120){\circle*{10}}

\put(100,80){\circle*{10}}
\put(120,80){\circle{10}}\put(140,80){\circle*{10}}
\put(80,80){\circle{10}}\put(60,80){\circle*{10}}

\put(100,140){\circle{10}}\put(120,140){\circle*{10}}\put(80,140){\circle*{10}}

\put(100,60){\circle{10}}\put(120,60){\circle*{10}}\put(80,60){\circle*{10}}

\put(100,40){\circle*{10}}\put(100,160){\circle*{10}}

\put(140,100){\oval(80,20)[t]}\put(60,100){\oval(80,20)[t]}
\put(100,60){\oval(30,80)[r]}\put(100,140){\oval(30,80)[r]}

\put(180,100){\circle{20}}\put(20,100){\circle{20}}
\put(100,180){\circle{20}}\put(100,20){\circle{20}}}      
\end{picture}
\caption[]{\small Renormalization scheme for $\nu=4$: 
the diagonal terms $f_{i,j}\equiv\eps_{i,j}\psi_{i,j}$
that form the block potential are painted black and show a
chess-board pattern. The sites $(n,m)$, $(n+4,m)$, $(n-4,m)$, $(n,m+4)$ and
$(n,m-4)$ that couple via a Laplace operator of distance $\nu$ are symbolized by
larger circles.
}
\label{bi:sketch2}
\end{figure}}
This procedure can be continued. By replacing again the site indices $(n,m)$ 
of Eq.~(\ref{anderson4}) by $(n+1,m)$, $(n-1,m)$, $(n,m+1)$ and $(n,m-1)$
and following the same procedure as before, we arrive at block indices $\nu=4$.  
As long as the block length
is well below $\Lambda/2$, the Taylor expansion is legitimate and we arrive 
at higher and higher orders of the renormalization. The potential blocks form a 
chess-board pattern which is shown in Fig.~\ref{bi:sketch2} for the case of $\nu=4$. 
The renormalized Anderson equation of block length $\nu$ becomes
\begin{eqnarray}\label{andersonfinal}
&& \psi_{n+1,m}^{(\nu)} + \psi_{n-1,m}^{(\nu)} + \psi_{n,m+1}^{(\nu)} +
\psi^{(\nu)}_{n,m-1} - 
4 \psi^{(\nu)}_{n,m} \nonumber\\
&& = -\eps^{(\nu)}_{n,m}\,\psi^{(\nu)}_{n,m}
+\nu^2\,(E-4)\,\psi^{(\nu)}_{n,m},
\end{eqnarray}
where $\psi^{(\nu)}_{n,m}$ is the smoothed wavefunction of a block of 
length $\nu$ and
\beq
\eps^{(\nu)}_{n,m} \equiv\sum_{i^2+j^2<\nu^2,i+j\,\rm{odd}} \eps_{i,j}
\eeq
with the sum running over all pairs of $i$ and $j$ with 
$i$ even, $j$ odd and vice versa (chess-board pattern) in a distance
$i^2+j^2<\nu^2$ from the site index $(n,m)$.

\section{The scaling ansatz: theory and numerical simulations}
Now, the renormalization approach is complete and
we use it to derive a scaling theory for the localization volume $V$. 
Naturally, the form of the wavefunction $\psi\sim\exp[-(r/\lambda)^\Phi]$
does not depend on the arbitrary subdivision of the lattice into blocks.
Nevertheless, by applying the renormalization approach over a certain range of
block lengths, we gain information about $V$.

The following derivation applies at $E=4$, where $\Lambda$ diverges and
the block form is legitimate for any block size between $1$ and infinity.
At $E=4$, the only quantities that enter into the right-hand 
side of Eq.~(\ref{anderson}) are the potentials $\eps_{i,j}$. Accordingly
$V$, which is an average quantity over many lattice realizations, 
can only depend on the different moments of them (the first moment $\eav$ 
being zero). As in $d=1$ we presume a power-law behavior,
\beq\label{power}
V \sim\var ^{-\alpha}.
\eeq
To derive the exponent $\alpha$, we apply the block transformation described above
separately to both sides of Eq.~(\ref{power}). The left-hand side, $V$, 
is a volume and therefore simply rescaled by a factor of $\nu^2$,
\beq\label{lamtrans}
V\to V_\nu \sim  \fr{V}{\nu^2}.
\eeq

The right-hand side of Eq.~(\ref{power}) is determined by random walk
theory. If we want to transform $\var$ into $\var_\nu$, we must first summarize 
over all $\nu^2$ potentials $\eps_{i,j}$ of one block and then calculate 
the variance over many different blocks. This is equivalent to 
calculating the mean square 
displacement of a random walk of $\nu^2$ steps \cite{Bunde2},
\beq\label{vartrans}
\var\to\var_\nu = \left\langle\left(\sum_{i=1}^{\nu^2} \eps_i\right)^2\right\rangle
\sim \nu^2\var.
\eeq
Transforming Eq.~(\ref{power}) by Eqs.~(\ref{lamtrans}) and 
(\ref{vartrans}) we find
\beq\label{skalend}
\fr{V}{\nu^2} = \nu^{-2\alpha}\var^{-\alpha}.
\eeq
As the last step, we must take into account that the block length $\nu$ is 
arbitrary for $\Lambda\to\infty$ and Eq.~(\ref{skalend}) must therefore not 
depend on $\nu$. This determines the exponent $\alpha$ and we finally find
\beq\label{skal0}
\alpha=1\quad\rm{and}\quad V\sim\var^{-1}\quad\rm{for}\quad E=4.
\eeq

\unitlength 1.85mm
\vspace*{0mm}
{
\begin{figure}
\begin{picture}(40,36)
\def\epsfsize#1#2{0.4#1}
\put(0,0){\epsfbox{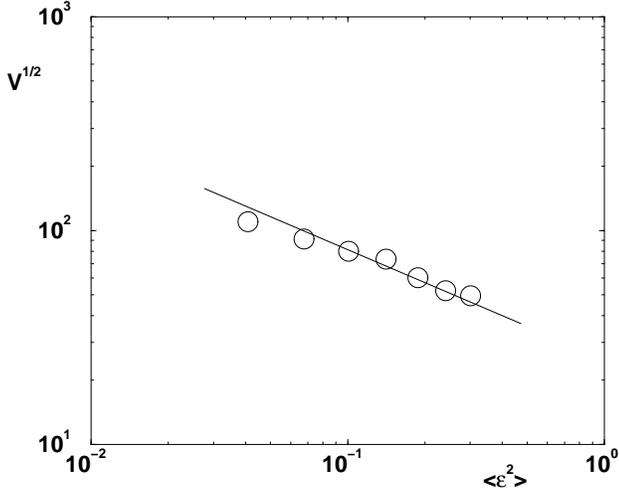}}
\end{picture}
\caption[]{\small The effective localization length $V^{1/2}$ in
the $2D$-Anderson 
model at the band edge is plotted versus the variance $\var$ of the site 
potentials in a double-logarithmic plot. $V$ was calculated numerically
for lattices of size $500\times 500$ with Dirichlet boundary conditions
and averaged over $40$ systems. The line of slope $-1/2$ is a guide to
the eye and shows the theoretical behavior. Finite size effects occur for
small values of $\var$ (large values of $V^{1/2}$).
}
\label{bi:skal0}
\end{figure}}
In order to test Eq.~(\ref{skal0}), the eigenfunctions of systems of
size $500\times 500$ with varying variances $\var$ have been calculated by 
the Lanczos algorithm. The different $V$ have been determined using
Eq.~(\ref{ipr}). For each $\var$, we took the average over $40$ systems
and calculated the eigenfunctions in a small energy interval of $E=4\pm
0.0002$. The results are shown in Fig.~\ref{bi:skal0}, where 
$V$ is plotted versus $\var$ in 
a double-logarithmic way. The line of slope $-1/2$ is a guide to
the eye and represents the result of the scaling theory [see Eq.~(\ref{skal0})].
Apart from slight finite-size effects for small $\var$ (and therefore 
large $V^{1/2}$) it agrees very well with the numerical results.

The scaling theory can be extended to energies in some (small) distance
from the band edge, where $\Lambda$ is still large enough to perform the 
renormalization scheme over many steps. In Eq.~(\ref{andersonfinal}), $(4-E)$ 
is rescaled with $\nu^2$. Accordingly, we have (cf. Eqs.~(\ref{lamtrans}) and
(\ref{vartrans}))
\beq \label{homo}      
V_{\nu=1}\left((4-\ve E\ve),\var\right) \sim \nu^2\,      
V_{\nu}\left(\nu^{2}(4-\ve E\ve),\nu^{2}\var\right).      
\eeq      
Equation~(\ref{homo}) is a generalized homogeneity relation.
This means that the form of $V$ remains unchanged when both, $\var$ 
and $4-\ve E\ve$, are rescaled according to the
renormalization theory. Thus, $V$ does not depend on both 
quantities separately, but only on a suitable combination of them. 

The scaling form of $V$ can now be derived by standard techniques.
Choosing $\nu=\var^{-1/2}$, (which is permitted for large $\var$ even if
$\Lambda$ is not infinite) we find
\begin{equation} \label{skalfkt}      
V((4-\ve E\ve),\var) \sim    
\var^{-1}\,g\left(\fr{4-\ve E\ve}{\var}\right)   
\end{equation}      
with the scaling function $g(x)$ and the argument 
\beq\label{skalarg}
x=\fr{4-\ve E\ve}{\var}. 
\eeq
For $\ve E\ve=4$, Eq.~(\ref{skalfkt}) must reduce to Eq.~(\ref{skal0}), 
yielding $g(0)=1$ for $x=0$. For small values of $4-\ve E\ve$ (large $\Lambda$)
or large values of $\var$ (small $V$), $\sqrt{V}\ll\Lambda$ and $x\ll 1$.
In this case, the effective localization length $\sqrt{V}$ is smaller 
than $\Lambda$, the system behaves as if $\Lambda$ were infinite and $g(x)$
should therefore be a constant function.
For $x\gg 1$, the maximum block size becomes smaller and smaller, so that 
gradually, the scaling theory must
break down. However, as in $d=1$, an intermediate range may exist, where
$g(x)$ still shows a power-law behavior.

\unitlength 1.85mm
\vspace*{0mm}
{
\begin{figure}
\begin{picture}(40,40)
\def\epsfsize#1#2{0.4#1}
\put(2,0){\epsfbox{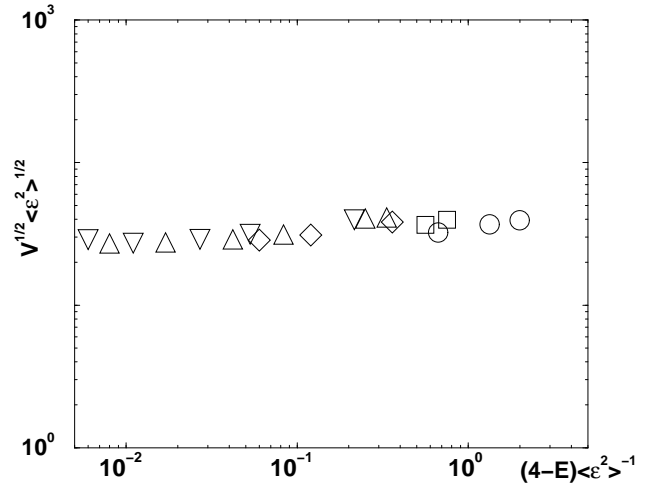}}
\end{picture}
\caption[]{\small As a test of Eq.~(\ref{skalfkt}), $(V\var)^{1/2}$ is plotted 
versus the argument $x\equiv (4-E)/\var$ 
for different disorder widths $w=0.6$ (circles), $w=0.8$ (squares), $w=1.0$ 
(diamonds), $w=1.2$ (triangles up) and $w=1.5$ (triangles down) with $\var=w^2/12$
and for different values for $4-E$ between $0.001$ and $0.06$. 
The average was again taken over $40$ systems of size $500\times 500$.}
\label{bi:skal1}
\end{figure}}
In order to test Eq.~(\ref{skalfkt}), we have plotted $(V\,\var)^{1/2}$
as a function of $(4-E)/\var$ for different disorder widths $w$ of the potentials,
$\eps_i\in[-w/2,w/2]$ with $\var=w^2/12$ and for different values of
$4-E$. The numerical simulations were again carried out on $500\times
500$ lattices and the average was performed over $40$ systems and in an energy
interval $[4-E-0.0002,4-E+0.0002]$ for different values of $4-E$ between 
$0.001$ and $0.06$. 
With decreasing values of $\var$, $V$ increases and
finite size effects occur. Additionally,
systems where $\var$ and $(4-E)$ are both small or both 
large, possess large error bars, i.e. large fluctuations between different 
values of $V$. In the case of small $\var$,
this also gives rise to finite size effects, because some very large values of 
$V$ are suppressed by the finite system size. So, we restrict 
ourselves to not too large values of $4-E$ and to
combinations, where such large fluctuations do not occur.

The results are shown in Fig.~4 and confirm the scaling ansatz (\ref{skalfkt})
very well. Different symbols that indicate different $\var$ fall onto the 
same universal curve. We can see that indeed $g(x)$ reaches a plateau, 
$g(x)\approx\rm{const}$ for small values of $x$ where 
$V$ is simply described by Eq.~(\ref{skal0}) (see above).
For large values $x\gg 1$, on the other hand, the scaling theory must break down, 
possibly after 
an intermediate range with a different power-law behavior of $g(x)$. It would 
be very interesting to investigate also this regime, 
However, large values of $x$ have not been calculated, because  -- due to the
increasing vaues of $\sqrt{V}$ -- we needed much 
larger system sizes for the simulations. This is currently not possible.

\section{Conclusive Remarks}
In summary, a renormalization scheme has been developed close to the band edges
(i.e. in the limit of large wavelengths) that analytically reduces
the Anderson equations (\ref{anderson}) into block form where
the block sizes may become arbitrarily large at the band edges. 
A scaling form for the localization volume $V$ has been derived
from this.
Contrary to former renormalization schemes \cite{renorm1,renorm2}, it does not 
involve sucessive recalculations of the matrix elements in each step, but
simply replaces Eq.~(\ref{anderson}) by Eq.~(\ref{andersonfinal}), where the 
off-diagonal elements are unchanged and the diagonal elements of arbitrary 
block size $\nu$ are directly related to the diagonal elements of the original 
system. The works of \cite{renorm1,renorm2} proposed the mobility edge in 
$d=3$ and scaling
laws for the conductivity and related quantities. Therefore, it will be very 
interesting to extend also the present theory to the three-dimensional Anderson 
model. However, as it is developed for energies close to the band edge 
it is for the moment not clear, if it can be applied to the vicinity of the
mobility edge, where comparisions to former renormalization theories can be
made.

As a last remark, we would like to note that
the regime $x\gg 1$ is also relevant to the vibrational
problem with unit spring constants and fluctuating masses $m_{n,m}=
\lan m\ran + \tilde m_{n,m}$, where $\lan m\ran$ describes the average
mass and $\tilde m_{n,m}$ the disorder of 
them. If we transform Eq.~(\ref{anderson}) according to 
\beq\label{transform}
4-E\to\mav\omega^2,\quad
\eps_{n,m}\to\tilde m_{n,m}\omega^2
\eeq
with the eigenfrequency
$\omega$ of the vibration, we find the vibrational equation 
\beq\label{schwing}
\fr{1}{m_{n,m}}\sum_{n',m'}(\psi_{n',m'} - \psi_{n,m}) =  -\omega^2 \psi_{n,m},
\eeq
with the sum going over all neighbors of the site $(n,m)$. Inserting the above
transformation into Eq.~(\ref{wavelenght}), we find for the wavelength in the 
vibrational case $\Lambda\stackrel{>}{\sim}(\mav^{1/2}\omega)^{-1}$. The limit of long 
wavelengths applies thus for $\omega^2<1/\mav$. Positive masses lead to
$\mvar<\mav^2$ and together with Eq.~(\ref{skalarg}) we finally arrive at $x>1$.

So, in the vibrational case, only the branch of higher values of the
scaling variable $x$ exists and it will be very interesting to investigate 
also this part. However, since this demands much larger 
system sizes (due to the increasing values of $V$), this should be done 
in the future.

\noindent {\bf Acknowledgements} I would like to thank
A. Bunde and J.W.~Kantelhardt for a careful reading of the manuscript
and interesting remarks.

\end{multicols}

\end{document}